\documentclass[pdflatex,sn-mathphys, iicol]{sn-jnl}
\theoremstyle{thmstyleone}%
\theoremstyle{thmstyletwo}%
\theoremstyle{thmstylethree}%
\usepackage{lineno}

\begin{document}

\title[Xu et al.]{Predicting binding motifs of complex adsorbates using machine learning with a physics-inspired graph representation}

\author[1,2]{\fnm{Wenbin} \sur{Xu}} 
\author[2]{\fnm{Karsten} \sur{Reuter}} 
\author*[3,4]{\fnm{Mie} \sur{Andersen}}\email{mie@phys.au.dk}

\affil[1]{\orgdiv{Chair for Theoretical Chemistry and Catalysis Research Center}, \orgname{Technische Universit{\"a}t M{\"u}nchen}, \orgaddress{\street{Lichtenbergstr. 4}, \city{Garching}, \postcode{D-85748}, \country{Germany}}}

\affil[2]{\orgname{Fritz-Haber-Institut der Max-Planck-Gesellschaft}, \orgaddress{\street{Faradayweg 4-6}, \city{Berlin}, \postcode{D-14195}, \country{Germany}}}

\affil*[3]{\orgdiv{Aarhus Institute of Advanced Studies,}
\orgname{Aarhus University}, \orgaddress{\street{H\o{}egh-Guldbergs Gade 6B}, \city{Aarhus C}, \postcode{8000}, \country{Denmark}}}

\affil*[4]{\orgdiv{Department of Physics and Astronomy - Center for Interstellar Catalysis,}
\orgname{Aarhus University}, \orgaddress{\street{Ny Munkegade 120}, \city{Aarhus C}, \postcode{8000}, \country{Denmark}}}

\abstract{Computational screening in heterogeneous catalysis relies increasingly on machine learning models for predicting key input parameters due to the high cost of computing these directly using first-principles methods. This becomes especially relevant when considering complex materials spaces, e.g.\ alloys, or complex reaction mechanisms with adsorbates that may exhibit bi- or higher-dentate adsorption motifs. Here we present a data-efficient approach to the prediction of binding motifs and associated adsorption enthalpies of complex adsorbates at transition metals (TMs) and their alloys based on a customized Wasserstein Weisfeiler-Lehman graph kernel and Gaussian Process Regression. The model shows good predictive performance, not only for the elemental TMs on which it was trained, but also for an alloy based on these TMs. Furthermore, incorporation of minimal new training data allows for predicting an out-of-domain TM. We believe the model may be useful in active learning approaches, for which we present an ensemble uncertainty estimation approach.}

\maketitle

Many surface catalytic reactions of vital importance to our society such as Fischer-Tropsch, methanol, or higher oxygenate synthesis have complex reaction mechanisms with numerous intermediates ranging from atoms and simple molecules to (possibly oxygenated) C1, C2 or larger fragments. It is well-known that modeling of these latter complex species at transition metal (TM) catalysts must account for their ability to exhibit a wide range of adsorption motifs, including mono-, bi- and higher-dentate adsorption modes.\cite{cao2018mechanistic,chang2021application, wang2020co2} Density-functional theory (DFT) with van der Waals corrections can, in principle, provide the energetics of such adsorption motifs at moderate cost and satisfactory accuracy.\cite{norskov2011density} Nevertheless, already the identification of the most stable adsorption motifs of adsorbates involved in ethanol synthesis at a simple monometallic catalyst such as Rh(111) is a formidable task,\cite{choi2009mechanism, michel2011c, filot2015firstprinciplesbased, gu2020automated} and the investigation of broader classes of materials such as TM alloys is generally out of reach due to the combinatorial explosion of possible active sites and adsorption motifs.

Machine learning (ML) models have already shown their potential for replacing expensive DFT calculations in order to tackle the screening of large materials spaces for accelerated catalyst discovery.\cite{tran2018active, noh2018active, wang2021infusing, andersen2019scaling, fung2021machine} However, most works so far have been limited in scope to the consideration of atoms or small molecules with mono-dentate adsorption motifs. For these simple species, models now routinely achieve the prediction of adsorption enthalpies with a root-mean-square-error (RMSE) around 0.1--0.2~eV, which is then comparable to the intrinsic DFT accuracy.\cite{andersen2019scaling, fung2021machine, back2019convolutional, gu2020practical} 

Unfortunately, most of these methods cannot easily be extended to complex adsorbates with bi- or higher-dentate adsorption motifs. One notable attempt to treat complex adsorbates is provided in the Open Catalyst Project where the direct prediction of relaxed adsorption enthalpies is achieved by incorporating a graph representation of the initial structure into a graph convolutional neural network.\cite{chanussot2021open} However, approaches that rely purely on connectivity and geometry-based features have revealed poor data efficiency and thus cannot be used without excessively large training databases.\cite{back2019convolutional,gu2020practical,chanussot2021open} Moreover, the predictive performance for complex adsorbates in the Open Catalyst Database is still below practical usefulness with a mean absolute error (MAE) for in-domain prediction around 0.6~eV.\cite{chanussot2021open}

In this work we develop and test a data-efficient, physics-inspired ML model applicable for both simple and complex adsorbates based on graph representation, the Wasserstein Weisfeiler-Lehman (WWL) graph kernel \cite{togninalli2019wasserstein}, and Gaussian Process Regression (GPR). We abbreviate the model WWL-GPR. For comparison, we show also results for predictions of simple and complex adsorbates using other popular, fundamentally different ML approaches that employ input in vector form instead of graph representation, namely the Sure Independence Screening and Sparsifying Operator (SISSO) approach \cite{ouyang2018sisso,ouyang2019simultaneous}, GPR with a radial basis function kernel (RBF-GPR) and eXtreme Gradient Boosting (XGBoost).\cite{chen2016xgboost} We train our ML models for complex adsorbates on a relatively small database (around 1700 data points) of DFT adsorption enthalpies calculated at the face-centered cubic (fcc) (211) and (111) facets of four TMs; copper (Cu), rhodium (Rh), palladium (Pd), and cobalt (Co). The chosen adsorbates and TMs are of interest for ethanol synthesis.\cite{medford2014activity,schumann2018selectivity}

Compared to the Open Catalyst Dataset, our dataset is smaller by about a factor of 300, covers less diverse surfaces and adsorbates, but exhibits a much denser sampling of diverse adsorption motifs for each catalyst / adsorbate combination considered. 
More importantly, we do not rely on graph representation alone, but augment it with node attributes representing physically motivated properties, e.g.\ $d$-band moments (surfaces), highest-occupied and lowest-unoccupied molecular orbital (HOMO/LUMO) energy levels (adsorbate molecules) and features of the local geometry, all derived from either the clean surfaces or the adsorbates in the gas phase. The model achieves an in-domain prediction of adsorption enthalpies with a RMSE of about 0.2~eV and also shows good extrapolative performance for two test cases; bimetallic alloys made from elements present in our training data and out-of-domain elements, the latter however only after incorporation of adsorption enthalpies of atomic species on the new element into the training database. Finally, we show that data points with large prediction errors can be quite reliably captured from an ensemble uncertainty estimation approach. 

\section*{Results}\label{sec2}

\subsection*{WWL-GPR model}\label{subsec2}
The ML task in our work is to directly predict the relaxed adsorption enthalpies corresponding to a range of possible adsorption motifs represented as graphs. Thereby, for a given surface/adsorbate combination of interest, we obtain a spectrum of possible adsorption energies ranging from the most stable to meta-stable adsorption motifs. Microkinetic models used in catalyst screening often employ only the most stable adsorption energy obtained as input, however, distinct adsorption sites with less favorable adsorption energies could be included as well.\cite{deimel2020active} Our task is thus quite similar to the task denoted as IS2RE (initial state to relaxed energy) in the Open Catalyst Project, however, we do not directly use the initial state geometry, but only its graph representation. We note that an entirely different approach to this task is to train a ML interatomic potential \cite{deringer2019machine, gasteiger_gemnet_2021} to relax the initial structure and thereby predict both the relaxed structure and adsorption enthalpy -- such approaches are however not a topic of this work.

\begin{figure*}[htp]
\centering
\includegraphics[width = 1 \textwidth]{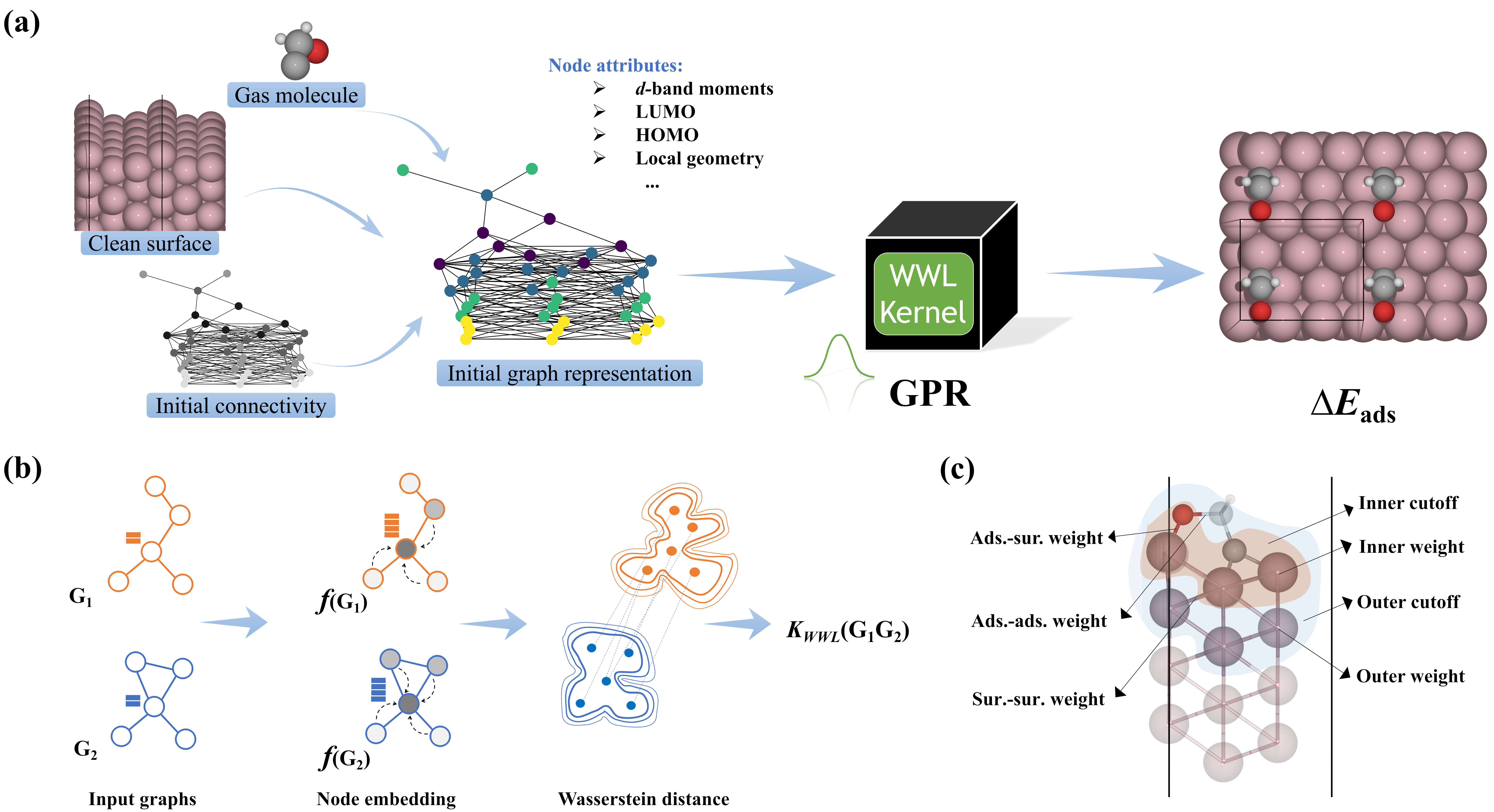}
\caption{\label{fig:Schematic} Schematic illustration of the WWL-GPR model. (a) The adsorption enthalpy of the relaxed structure, $\Delta E_{\rm ads}$, is predicted from a graph representation of the initial structure with node attributes computed from the gas-phase molecule and clean surface. The graph similarity is calculated from the WWL graph kernel and input to a GPR model. (b) The similarity of two input graphs in the WWL kernel, $K_{\rm WWL}$, is calculated by first generating node embeddings and then computing the Wasserstein distance between their distributions. (c) Surface adsorption motivated hyperparameters incorporated into the WWL kernel.}
\end{figure*}

Fig.\ \ref{fig:Schematic}(a) depicts a schematic of our physics-inspired WWL-GPR model. We rely on graph representation, which is a versatile method for representing isolated molecules \cite{wen2021bondnet,tang2019prediction}, crystal structures,\cite{xie2018crystal} or the combined surface/adsorbate system,\cite{montoya2017highthroughput,boes2019graph,deshpande2020graph} in which every atom in the structure is a node with edges representing chemical bonds to neighboring atoms. Graph representation can be used in connection with neural networks,\cite{chanussot2021open,gu2020practical, back2019convolutional} which generally requires very large training databases. Since we are here interested in developing a data-efficient method, we focus on a kernel-based method (GPR) in connection with a customized version of the recently developed WWL graph kernel.\cite{togninalli2019wasserstein} Fig.\ \ref{fig:Schematic}(b) illustrates the node embedding scheme, the calculation of the Wasserstein distance (distribution relationship) between the graphs, and the subsequent WWL graph kernel calculation. The WWL graph kernel allows for continuous node attributes, for which we use physically motivated electronic and geometric features calculated from the clean surface and isolated adsorbate. Finally, we incorporate into the WWL kernel some surface adsorption motivated hyperparameters to learn better representations, see Fig.\ \ref{fig:Schematic}(c); edge weights, which differentiate chemical bonds in the three classes adsorbate-adsorbate, surface-surface and adsorbate-surface, as well as inner and outer cutoffs and weights. The latter are used during the computation of the Wasserstein distance to emphasize the importance of various atomic shells around the active site for the adsorption energy prediction. We note that attention algorithms widely used in neural networks serve a similar purpose.\cite{gu2020practical}


\subsection*{Prediction of simple adsorbates}\label{subsec3}
\begin{figure*}[htp]
     \centering
     \includegraphics[width = 1 \textwidth]{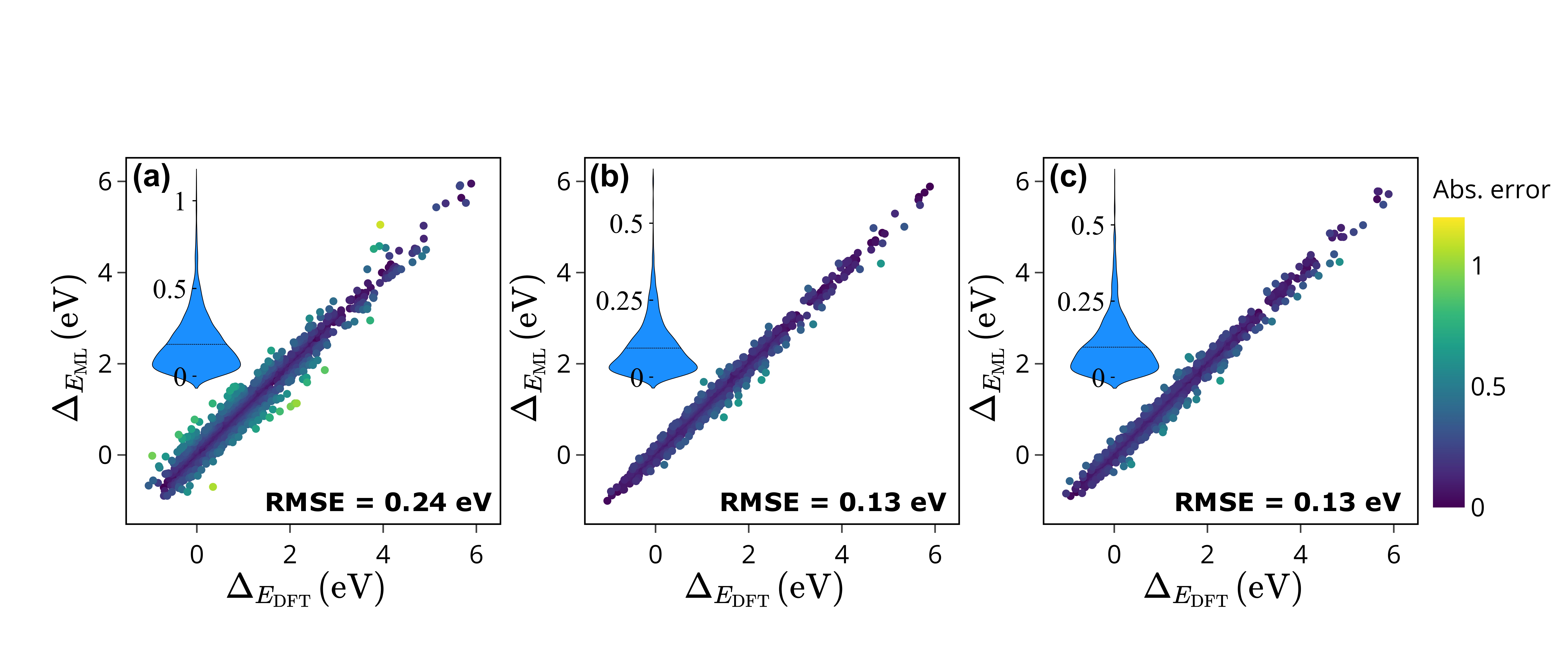}
     \caption{\label{fig:small_ads_prediction} Parity plot of DFT-calculated vs.\ ML-predicted adsorption enthalpies from the combined validation set in 5-fold cross validation using (a) SISSO, (b) RBF-GPR, (c) WWL-GPR for the simple adsorbates database. The violin plots in the insets illustrate the absolute (Abs.) error distributions (in eV), and the internal dashed line in the plots marks the mean absolute error.}
\end{figure*}

We begin by evaluating the performance of the ML models for predicting a database of simple adsorbates with mono-dentate adsorption motifs (see Methods section). We carry out 5-fold cross validation, that is, the database is shuffled and partitioned into five equal-sized subsamples stratified by adsorbates. The training is then carried out based on four of the subsamples while retaining the fifth subsample for validation. This is repeated five times until all data points have been used once for validation. Fig.\ \ref{fig:small_ads_prediction} shows the resulting parity plot of DFT-calculated against ML-predicted adsorption enthalpies for the combined validation set from the five folds as well as violin plots of the absolute error distributions for SISSO and the GPR models. It should be noted here that the SISSO results are obtained using similar hyperparameters as in our previous work \cite{andersen2019scaling, xu2020data} (eight-dimensional rung three descriptor). In principle, we would expect a better performance than the here presented RMSE of 0.24~eV for even more complex models, see Supplementary Fig.\ 5. However, the identification of more complex models is computationally intractable with the SISSO method. Rather than raw performance, the merit of the SISSO approach is that the identified descriptors are (somewhat simple) analytical functions of the features, which are thus easier to interpret than black-box ML models. We also note that the reason for the different performance of the descriptors identified in the present work compared to our previous work is that here we train a single model on the entire database (single-task learning) in order to be able to make a direct comparison to the GPR models, whereas in our previous work separate fitting coefficients were used for each adsorbate (multi-task learning). More information about the identified SISSO descriptors is provided in Supplementary Section 3.2.

For the GPR models the model complexity can be more easily tailored and after optimization of the relevant hyperparameters (see Supplementary Section 3.5 and Supplementary Table 9) we obtain a RMSE of 0.13~eV independently of whether we use vector input (RBF-GPR) or graph representation (WWL-GPR). Also the maximum absolute error (maxAE) decreases from 1.11~eV (SISSO) to around 0.60~eV in the GPR models. Based on the similar performance of the two GPR models, we can conclude that there is no added value from employing graph representation for the simple adsorbates. The reactivity is apparently already captured by the averaged surface atom features and the adsorbate-specific features used in the RBF-GPR model. 

Finally, the XGBoost method represents an ensemble-based ML method based on decision trees and gradient boosting, where trees are added one at a time to improve on the residuals of the previous model.\cite{chen2016xgboost} Here, we find that it performs similarly to the GPR models (see Supplementary Fig.\ 8) with a RMSE of 0.12~eV. On the basis of this similar performance of state-of-the-art methods, we believe that we are at the limit of the ML accuracy achievable for simple adsorbates with the available data set and feature representation.

\subsection*{Prediction of complex adsorbates}\label{subsec4}

\begin{figure*}[htp]
\centering
\includegraphics[width = 1 \textwidth]{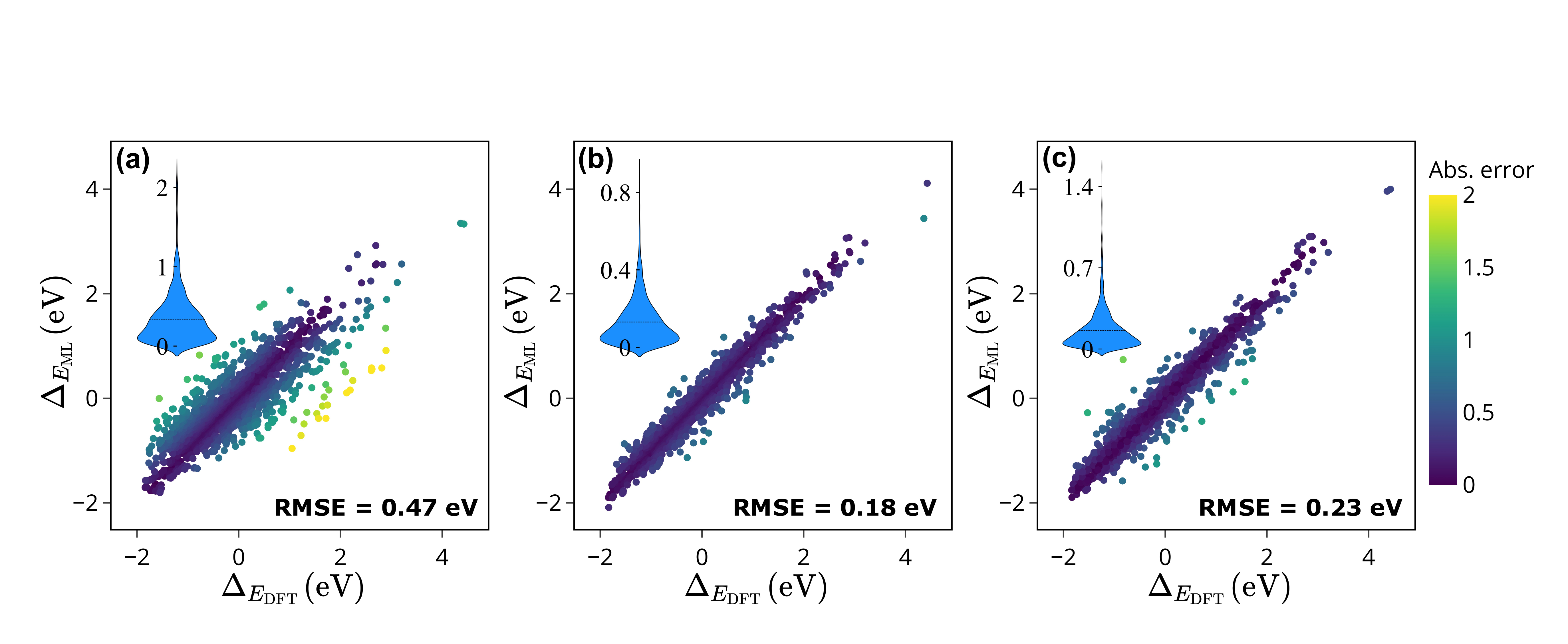}
\caption{\label{fig:complex_ads_prediction} Parity plot of DFT-calculated vs.\ ML-predicted adsorption enthalpies from the combined validation set in 5-fold cross validation using (a) RBF-GPR, (b) WWL-GPR and (c) XGBoost for the complex adsorbates database. The violin plots in the insets illustrate the absolute (Abs.) error distributions (in eV), and the internal dashed line in the plots marks the mean absolute error.
}
\end{figure*}

We next turn to a database of complex adsorbates with 41 different adsorbates in mono-, bi-, and higher-dentate adsorption motifs on surfaces of Cu, Co, Pd and Rh (see Methods section). Since we already concluded in the preceding section on simple adsorbates that single-task SISSO is not competitive in terms of performance, we focus here only on the GPR models and XGBoost. The 5-fold cross validation results presented in Fig.\ \ref{fig:complex_ads_prediction} show that for this more challenging database the graph-based WWL-GPR model has a superior performance (RMSE of 0.18~eV) compared to RBF-GPR (RMSE of 0.47~eV). Also the maxAE decreases from 2.23~eV (RBF-GPR) to 0.92~eV in the WWL-GPR model. The XGBoost method clearly outperforms RBF-GPR with a RMSE of 0.23~eV, which is possibly related to the advantages of its ensemble-based approach. However, it is still inferior to WWL-GPR. We attribute this to the importance of the graph representation for complex adsorbates, which is present in the WWL-GPR model but missing in the vector-based models. A learning curve for the WWL-GPR model is presented in Supplementary Fig.\ 6, which shows that an RMSE of 0.3~eV can be achieved by only training on 30\% of the database ($\sim$ 500 data points) and a RMSE of 0.2~eV is achieved at 70\% of the database ($\sim$ 1200 data points). A visualization of the prediction accuracy for adsorption motifs of one selected adsorbate (CHCO) on one selected surface (Cu(211)) is given in Supplementary Fig.\ 7.

\begin{figure*}[htp]
\centering
\includegraphics[width = 1 \textwidth]{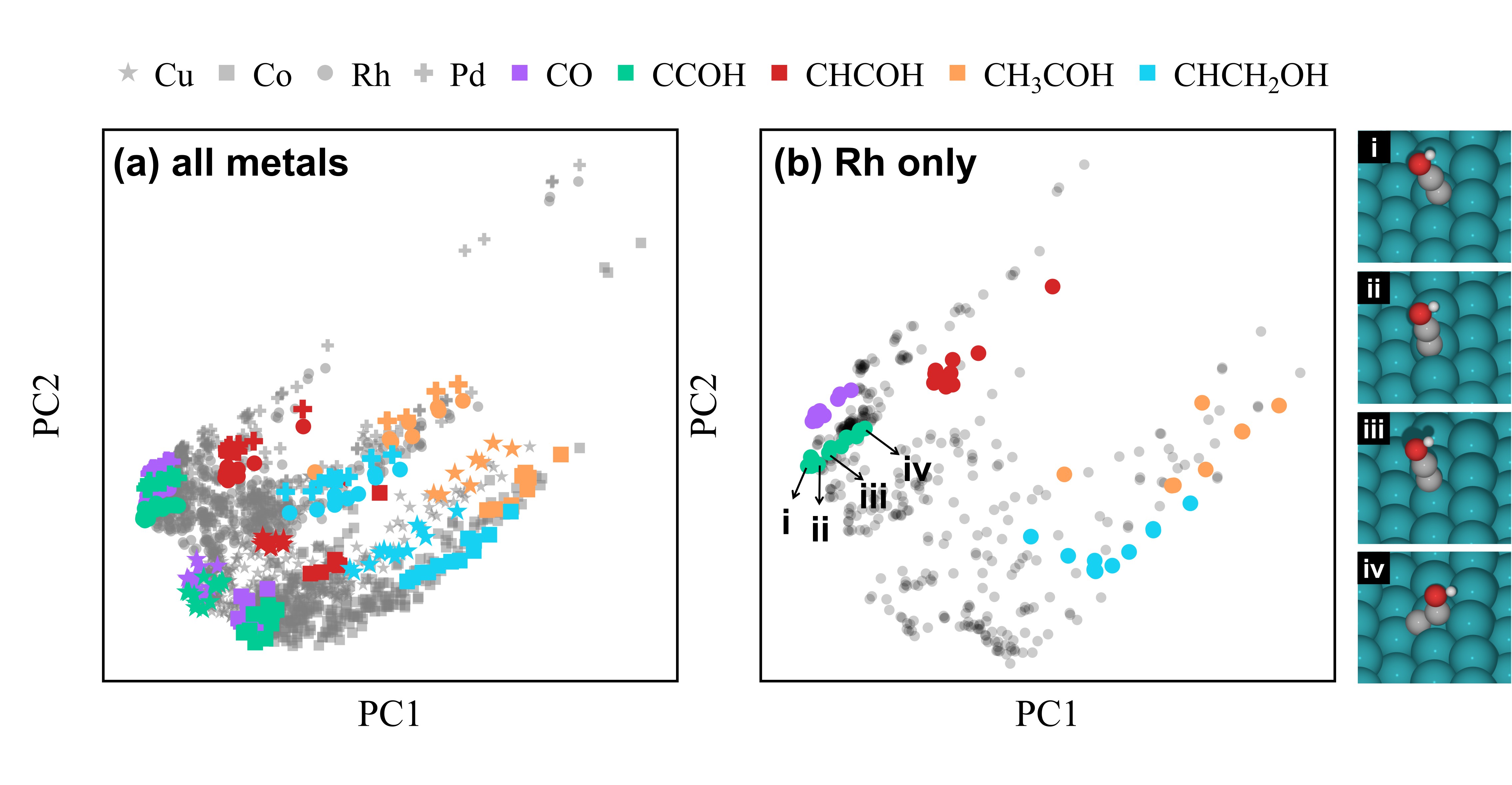}
\caption{\label{fig:kpca} Two-dimensional kernel principal component (PC) analysis plots for the WWL-GPR model with hyperparameters optimized for interpolation for (a) all metals and (b) Rh surfaces only. The locations and structures of selected adsorbates are highlighted. The Rh, C, O and H atoms are shown in green, grey, red and white, respectively.}
\end{figure*}

In order to visualize what trends the WWL-GPR model has identified in the complex adsorbates database, we present in Fig.\ \ref{fig:kpca} a kernel principal component analysis (KPCA), which is a non-linear dimensionality reduction technique. Specifically, we here present the two dimensions that explain the largest fraction of the variance. Points that are close together in this space are similar in the feature space. The analysis of the entire complex adsorbate database in Fig.\ \ref{fig:kpca}(a) shows that the different metals are distinguished as parallel clusters, where for each cluster there is a similar distribution of sub-clusters containing the individual adsorbates. In Fig.\ \ref{fig:kpca}(b) the same analysis is presented for only one metal (here Rh, but similar results are obtained for the other TMs). Again, the different adsorbates form clusters, where each point in a cluster corresponds to a separate adsorption motif of the adsorbate. A similar clustering cannot be observed in KPCA plots for the RBF-GPR model, see Supplementary Fig.\ 9, which is probably related to the fact that this model does not have any structural information about the different adsorbates and their associated adsorption motifs due to the lack of graph representation.

Having established the excellent interpolation performance of the WWL-GPR model, we next assess the predictive performance of the model for extrapolation tasks concerning data that are dissimilar to those in the training database, i.e.\ out-of-domain prediction. This is highly important for the practical application of the model to catalyst screening. The two tasks we consider are a) predictions for a bimetallic catalyst, i.e.\ an alloy of elemental metals present in our database, and b) predictions for an out-of-domain element when merely incorporating adsorption enthalpies of atomic species (C, H, and O) at the new element into the database. For these tasks we selected 8 adsorbates spanning both atomic species and larger molecules, and including some with bi-dentate adsorption motifs (see Supplementary Table 4). 

Since it has previously been emphasized in the literature that in extrapolative, data-poor regimes, a careful choice of regularization can substantially improve the robustness of a model \cite{rupp2015machine,deringer2021gaussian}, we re-optimized the hyperparameters for the extrapolation tasks. Specifically, we used data for one bimetallic alloy (CuCo) and one out-of-domain element, platinum (Pt), to optimize new hyperparameters by minimizing the loss function (RMSE$_{\rm interpolation}$ + $2*$RMSE$_{\rm extrapolation}$), where RMSE$_{\rm interpolation}$ is the RMSE for the original complex adsorbates database (including atomic adsorption enthalpies for Pt) and RMSE$_{\rm extrapolation}$ is the RMSE of the data set for CuCo and Pt (for Pt only the complex adsorbates). Since the aim is to find hyperparameters well suited for extrapolation, this latter task was given a higher weight (two) in the loss function than the weight of the interpolation task (one). Comparing the hyperparameters obtained previously for the complex adsorbates database (base case in Supplementary Table 10) with the new hyperparameters optimized specifically for the extrapolation tasks (base case in Supplementary Table 11), we see that indeed both the length scale and the regularization term increase for the extrapolation tasks, resulting in a smoother ML model, which is consistent with the previous literature observations. The obtained RMSEs for the new hyperparameters are 0.25~eV for interpolation within the complex adsorbate database, 0.23~eV for the CuCo alloy and 0.30~eV for Pt. Finally, we carry out a true extrapolation test to assess if the new hyperparameters would also be accurate for yet another bimetallic alloy (here we chose PdRh) and yet another out-of-domain element (here we chose Ru), see Supplementary Table 4. Indeed, we can obtain a very good extrapolation performance with a RMSE of 0.23~eV for PdRh and also 0.23~eV for Ru. We note here that apparently Ru is easier to extrapolate to than Pt (based on the lower RMSE obtained), which signifies that it must somehow be more similar to the elements present in the complex alloys database. Also, our results show that an out-of-domain element is generally harder to predict than an alloy of known elements, even when incorporating some minimal information about the unknown element into the training database through the atomic adsorption enthalpies. We would expect the performance for Pt to improve if more adsorbates were added to the training database.


\subsection*{Uncertainty quantification}\label{subsec5}

Up till now we have demonstrated that our WWL-GPR model can be applied with RMSE around 0.2--0.3~eV to flat and stepped metal and bimetallic catalysts, as long as some (at least minimal) training data involving the considered elemental metals are provided. However, apart from the average RMSE to expect, it is also useful to be able to directly assess the expected uncertainty on a single predicted data point. For example, uncertainty quantification (UQ) combined with sensitivity analysis of microkinetic models \cite{bruix2019first,meskine2009examination, medford2014assessing} can be used to assess error propagation and the extent to which conclusions drawn from a model are robust to input parameter uncertainty.\cite{sutton2016effects,dopking2017error} Furthermore, UQ is used in active learning approaches, where the training database is iteratively updated through selected DFT calculations, e.g., of data points with a high estimated uncertainty.\cite{flores2020active, kunkel2021active}

In view of these applications, we are here primarily interested in the extent to which a high estimated uncertainty correlates with a high actual error of the model predictions. To assess this point we use a random 80\%/20\% training/test split of the complex adsorbates database stratified by adsorbate. We compare the intrinsic UQ provided in a single GPR model trained on the training set through the standard deviation (SD) of the posterior distribution to the UQ provided by the SD of an ensemble (100 in total) of GPR models with fixed hyperparameters optimized for interpolation. The latter are constructed through bootstrapping of the training set, i.e.\ data points are drawn randomly with replacement. Note that the added computational cost of establishing the ensemble model is negligible since we use a fixed training/test split, and since the kernel between the training and test set only needs to be computed once.

\begin{figure*}[htp]
\centering
\includegraphics[width = 1 \textwidth]{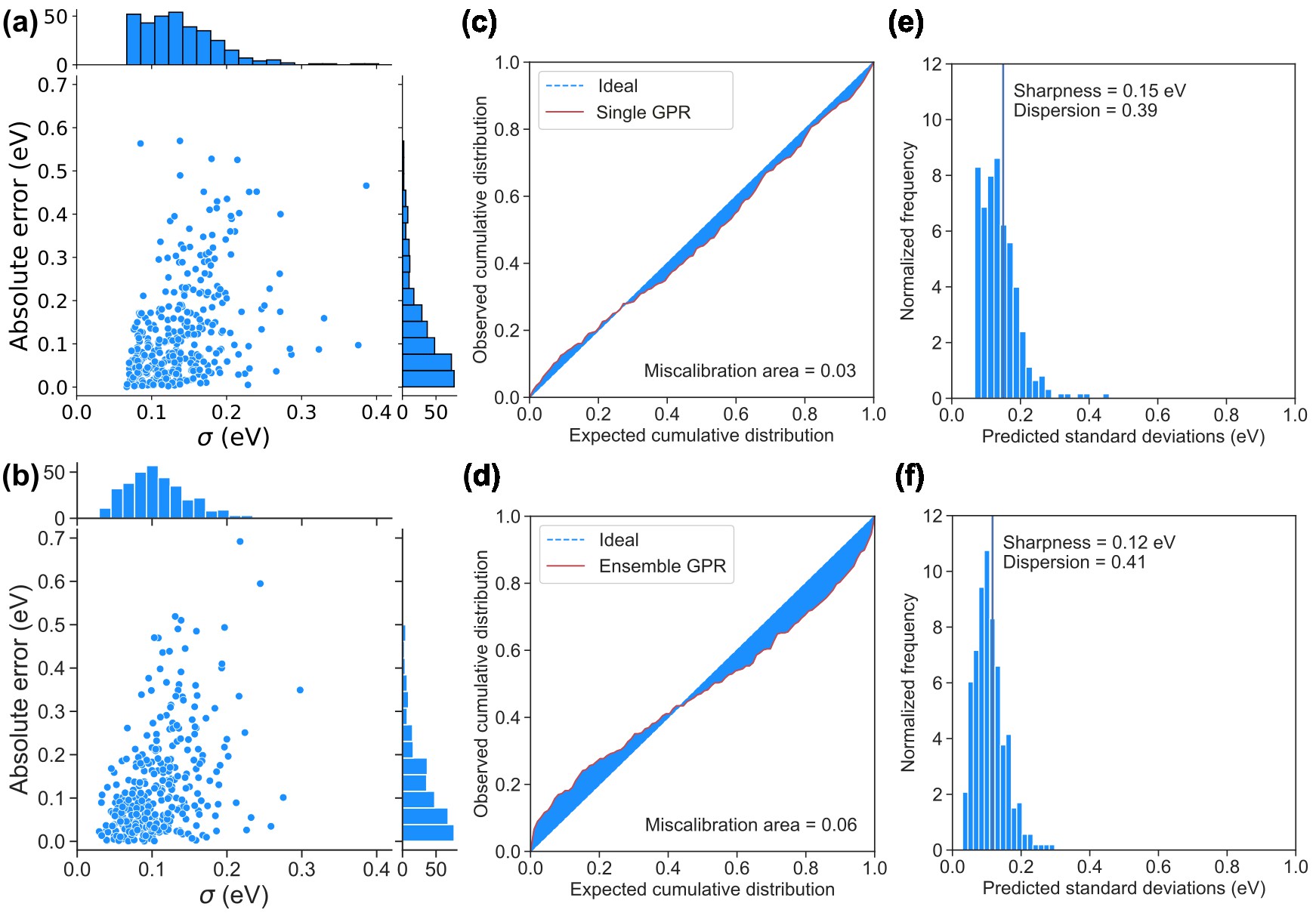}
\caption{\label{fig:uncerainty} The uncertainty vs. absolute error, calibration curve, and distribution plot of the predicted standard deviations for (a), (c) and (e) the single GPR model and (b), (d) and (f) the ensemble GPR model, respectively.}
\end{figure*}

As expected, the prediction accuracy obtained from the single and the ensemble model is almost identical (RMSE of 0.17~eV versus 0.18~eV, respectively). A plot of estimated uncertainties versus absolute prediction errors of the two models is presented in Fig.\ \ref{fig:uncerainty}(a) and (b). For comparison, we show also in Fig.\ \ref{fig:uncerainty}(c-f) some distribution-based measures of the quality of a UQ that have recently been discussed in the literature, i.e.\ calibration, sharpness, and dispersion.\cite{tran2020methods} A useful UQ method should have a small miscalibration area (a good match between the expected and observed cumulative error distribution), a small sharpness value (small error estimates) and a large dispersion value (disperse error estimates). For these latter quantities the performances of the single and ensemble models are quite similar with the single model having a slightly better calibration and the ensemble model having a slightly better sharpness and dispersion. However, with our primary interest being active learning, it is much more intriguing to see that the ensemble model does a better job than the single model at assigning a high uncertainty to data points with a high actual prediction error. In particular, the group of points with an estimated uncertainty higher than ~0.2~eV in the ensemble model includes the largest prediction errors, whereas this is not the case for the single model, which actually assigns a quite low uncertainty to some of the largest prediction errors. We therefore conclude that the ensemble model is best suited for active learning approaches. We note here that we do not expect any quantitative match between the absolute error and the uncertainty in Fig.\ \ref{fig:uncerainty}, partly because these are not directly comparable quantities (one is a SD and the other an absolute error), and partly because it has been shown that specific calibration measures \cite{palmer2022calibration,kuleshov2018accurate} are required for quantitatively accurate UQ in both single and ensemble GPR models. 


\section*{Discussion}\label{sec3}
We begin by discussing the origin of the superior performance of the WWL-GPR model over the vector-based RBF-GPR and XGBoost models. First of all, we note that it is not surprising that for complex adsorbates, simply accounting for the surface and adsorbate in terms of features averaged over the atoms directly involved in the bonding as well as global features of the adsorbate (e.g.\ HOMO/LUMO levels) and clean surface (e.g.\ work function) as done in the vector-based models is insufficient. In contrast, the graph representation provides direct access to structural information about the system, i.e.\ the number and types of atoms in the adsorbate and how these atoms connect to each other and to the surface, possibly in complex bi- or higher-dentate adsorption motifs. Atom-specific features related to the local electronic or geometric structure can be directly used as node attributes, e.g.\ through SOAP descriptors, and we can introduce surface adsorption motivated hyperparameters as discussed above and illustrated in Fig.\ \ref{fig:Schematic}(c). The main remaining limitation of our model is that it cannot be expected to handle cases where the adsorbate dissociates or the surface reconstructs upon the adsorption event, since it -- in contrast to ML force fields -- does not predict the entire potential energy surface of the system but only discrete minima corresponding to adsorbed states. Furthermore, it relies on user-specified features, which would have to be adjusted for the consideration of other materials classes, e.g., metal oxides.\cite{xu2020data}

Based on the demonstrated extrapolation performance, we trust that our WWL-GPR model could be useful for catalyst screening purposes, e.g.\ for exploring reactions with complex adsorbates on alloy surfaces. Here the complexity encountered from the many possible adsorption motifs of each adsorbate on each type of alloy surface makes direct DFT investigations computationally intractable, while reliable ML force fields or density-functional tight-binding methods for the simultaneous treatment of many different adsorbates and/or alloy surfaces are still difficult to obtain.\cite{chanussot2021open,chang2021application}

We envision that it could be particularly interesting to apply our model in the context of an active learning strategy, where the training database is iteratively expanded towards catalytically interesting and/or previously poorly explored regions of the catalyst space. Key advantages of our data-efficient GPR model in this regard are the low training cost (compared to, e.g., deep neural networks) and the demonstrated UQ.

For active learning purposes, we also recommend to use the model with different hyperparameter settings depending on the exploitative or explorative nature of the task at hand. Specifically, we can confirm previous literature reports that hyperparameters characterized, among others, by larger length scale and regularization terms are beneficial for accurate predictions in data-poor regions of the catalyst space. 

\section*{Methods}\label{sec4}

\subsection*{DFT databases}\label{subsec41}
The ML models are trained and tested on two different databases termed 'simple adsorbates' and 'complex adsorbates'. The database of simple adsorbates is taken from Refs.\ \cite{andersen2019scaling,deimel2020active}.
After a post-processing step the database contains 1422 data points and includes adsorption enthalpies of eight simple adsorbates with mono-dentate coordination; C, H, O, CO, OH, CH, CH$_2$ and CH$_3$. The considered surfaces include the fcc(100), fcc(110), fcc(111) and fcc(211) facets of pure Ni, Cu, Ru, Rh, Pd, Ag, Ir, Pt and Au, the body-centered cubic (bcc) (210) facet of Fe as well as the stepped hexagonal close-packed (hcp) (0001) facet of Co. For alloy catalysts, the database contains the adsorbates on the four single-atom alloys Ag@Cu, Pt@Rh, Pd@Ir, and Au@Ni (i.e.\ the single atom Ag, Pt, Pd or Au dispersed in the surface of another host metal) and the four AB bimetallic alloys AgPd, IrRu, PtRh, and AgAu.

The complex adsorbates database contains 1679 data points and includes 41 different small and large adsorbates involved in ethanol synthesis on fcc(111) and fcc(211) facets of Cu, Rh, Pd, and Co. Examples of complex adsorbates are CHCO, CCHOH, CH$_2$CH$_2$O and CH$_3$CH$_2$OH, and the full list of adsorbates is provided in Supplementary Table 3. Furthermore, selected adsorbates are calculated at the CuCo(111), PdRh(111), Pt(111), Ru(111), Pt(211) and Ru(211) surfaces for model testing purposes. The adsorbates contain up to nine atoms and cover mono-, bi-, and higher-dentate adsorption modes. The database is constructed using an automated workflow and DFT settings that are compatible with the simple adsorbates database (Quantum Espresso code \cite{Giannozzi_2009}, BEEF-vdW functional \cite{wellendorff2012density}). Further computational details and overviews of both databases are given below and in Supplementary Section 1.

\subsection*{Database construction and workflow}\label{subsec42}
The initial geometries of the surface/adsorbate systems are generated using the CatKit software.\cite{boes2019graph} CatKit employs a graph representation of the surface atoms to enumerate mono- and bi-dentate adsorption sites, where the latter are defined by a neighboring node-edge pair of the graph.
For each adsorbate, a manual tagging of the bonding atoms for mono- and bi-dentate adsorption motifs is required (see Supplementary Table 3).
CatKit then adds the adsorbates at the enumerated adsorption sites by employing some simple geometric procedures to produce good guesses for the angles and bond lengths in the system.
We note that CatKit obviously does not generate all possible adsorption motifs (which would be computationally intractable), but only those that are judged most plausible. This adds a human bias into the generation of the database. Furthermore, not all initial geometries generated are actually stable, but could transform into other structures during the DFT relaxation.

In order to overcome some of these limitations, we added the following steps to our computational workflow.
During the DFT relaxation, we monitor the graph representation of the system and assign it to the following four cases. (i) if the graph representation is unchanged, the data point is simply added to our database (32.3 \% of cases). (ii) if the structure transforms into another graph which is already covered in the CatKit-enumerated structures (28.4 \% of cases), only the calculation with the most favorable adsorption enthalpy is added to the database to avoid duplicates. (iii) if the structure transforms into a non-valid graph, i.e.\ a graph that is incompatible with our direct graph-based ML model (e.g.\ adsorbate dissociation, surface reconstruction) the calculation is discarded (23.4 \% of cases). (iv) if the structure transforms into a valid graph that was not enumerated by CatKit (15.9 \% of cases), the data point is added to the database with updated initial graph representation and the new adsorption motif is tested also for the other surfaces of interest. The latter case (iv) as well as large adsorbates whose initial adsorption motifs cannot be well controlled by CatKit are the source of all higher-dentate adsorption motifs in our database (see examples in Supplementary Fig.\ 3). Our workflow is implemented with AIIDA,\cite{huber2020aiida} which is a scalable computational infrastructure providing advanced automation to allow interfacing with external simulation software. In our case this entails customized python scripts \cite{wwlgpr2022} interfacing with CatKit, the Atomic Simulation Environment (ASE) software \cite{Hjorth_Larsen_2017} and the Quantum Espresso DFT code.

\subsection*{DFT computational details}\label{subsec43}
The simple adsorbates database used here is taken from Ref. \cite{deimel2020active} and \cite{andersen2019scaling}. For the DFT calculations of the complex adsorbates database, the following settings were used in full compliance with the simple adsorbates database. We used the Quantum ESPRESSO code \cite{Giannozzi_2009} with a plane-wave basis set, the Bayesian error estimation functional with van der Waals correlation (BEEF-vdW) \cite{wellendorff2012density} and ultrasoft pseudopotentials. Pseudopotentials for Cu, Rh, Pd, and Pt were generated using the 'atomic' code by A. Dal Corso (v.5.0.2 svn rev. 9415),\cite{dal2014pseudopotentials} for Co using the Vanderbilt code version 7.0.0,\cite{garrity2014pseudopotentials} and for Ru using the Vanderbilt code version 7.3.5. To relieve the interaction between the adsorbates, we modelled the fcc(211) slab in a $(3 \times 1)$ cell and the fcc(111) slab in a $(3 \times 3)$ cell. In both cases this corresponds to 9 atoms per atomic layer. The CuCo(111) and PdRh(111) alloy surfaces were modelled in a $(4 \times 2)$ cell and contain 16 atoms per layer.
We used a $(4 \times 4)$ k-point grid for the pure metal slabs and a $(3 \times 3)$ grid for the alloy slabs.
All slabs contained four atomic layers, where the bottom two layers were kept fixed in their bulk-truncated positions, while the top layers and the adsorbates were relaxed until the maximum force on each atom fell below 0.05~eV/$\textrm{\AA}$. See Supplementary Fig.\ 1 for images of the used slab geometries. All DFT calculations were carried out as periodic slab calculations employing a vacuum region of 20 $\text{\r{A}}$ perpendicular to the surface and a dipole correction. Spin polarization was taken into account for the calculations involving Co. The cutoff energy was set to 500 eV and 5000 eV for the orbitals, and the charge density, respectively, and a Fermi level smearing of 0.1 eV was used. The resulting adsorption enthalpies are formation energies referenced to gaseous CH$_3$OH, CO, and H$_2$O.

The features that require DFT calculations were obtained as follows. For the clean surfaces involved in both the simple and complex adsorbate databases, we first carried out a geometry relaxation as outlined above. The projected density of states (PDOS) was calculated using the smearing-free tetrahedron method and an energy spacing of 0.01~eV. We used a $(14 \times 14)$ k-point grid for the pure metal fcc and bcc slabs, a $(7 \times 21)$ grid for the SG225 fcc alloys, a $(14 \times 21)$ for the SG221 fcc alloys, a $(7 \times 14)$ grid for the Co hcp slab, and a $(7 \times 42)$ for the hcp alloy structures. For the CuCo(111) and PdRh(111) alloy surfaces involved in the extrapolation tasks we used a $(12 \times 12)$ k-point grid.

For the calculation of band moments, we integrated empty bands up to the energy above the Fermi level where the PDOS had fallen below a value of 0.01 {\AA}$^{-3}$eV$^{-1}$. The features involving the density of states at the Fermi level were calculated using a smearing of 0.1~eV in the PDOS calculation, and the PDOS was averaged over the interval $\pm$0.1 eV around the Fermi level. For the calculation of adsorbate-specific features, we carried out a structural optimization of the isolated adsorbate positioned in a cubic supercell with a side length of 15 {\AA}. We used a Fermi-level smearing of 0.01~eV and the Brillouin zone integration was performed using the Gamma point only

\subsection*{Further details on ML models}\label{subsec44}
The WWL-GPR model is compared to three other ML models (SISSO, RBF-GPR and XGBoost) that do not use graph representation, but input in vector form with features of the clean surface and of the isolated adsorbates. The features used in the vector-based models are specific to the surface, site or adsorbate considered, where site-specific features are calculated by averaging over the metal atoms to which the adsorbate coordinates (clean surface features) or the bonding atoms of the adsorbate (isolated adsorbate features). The WWL-GPR model also uses atom-specific features as node attributes, for example, electronic properties of individual surface atoms or features of the local geometry of the clean surface and isolated adsorbate through Smooth Overlap of Atomic Positions (SOAP) descriptors.\cite{bartok2013representing} All details about the features used in the compared ML models are provided in Supplementary Section 2.

Supplementary Section 3 provides more information about each of the models, including a more in-depth discussion of hyperparameters. When comparing the RBF-GPR and WWL-GPR model, it is interesting to note that while the WWL-GPR model finds that the optimal cutoff values are one node distance for both inner and outer cutoff for the simple adsorbates database (i.e.\ mostly the atoms directly involved in surface/adsorbate bonding are judged important), the optimal inner and outer cutoffs (weights) are one (0.60) and two (0.06) node distances, respectively, for the complex adsorbates database, see Supplementary Table 9 (i.e.\ also atoms neighboring the immediately bonding atoms are judged important, although with smaller weights). The effect of more distant atoms is not taken into account in the vector-based models, which then possibly relates to their decreased performance for complex adsorbates. Note also that during the node embedding scheme of the WWL graph kernel, the node attribute of every atom is updated with information about the node attributes of the neighboring atoms, see Supplementary Section 3.3.1.2. That is, even if weights beyond the outer cutoff are zero, the atoms there can still have a non-negligible influence on the kernel value.

It should be emphasized that the WWL-GPR model leverages only features from the initial guess geometry, specifically the graph connectivity, and electronic and geometric features calculated from the clean surface and isolated adsorbate. From a computational screening point of view this is essential for keeping the computational cost of model predictions low.
The computationally most intensive part of the model prediction is the DFT calculation of the clean surface to obtain the node attributes (e.g.\ $d$-band moments) for the surface atoms. However, given that we target 41 different adsorbates in various possible adsorption motifs for each surface, this is still a low cost per ML prediction.

For SISSO, we previously used an approach to target simple adsorbates where the free parameters of the identified models were fitted to each adsorbate separately.\cite{andersen2019scaling, xu2020data} A similar approach has been taken in most other works targeting simple adsorbates.\cite{back2019convolutional, Esterhuizen2020, wang2021infusing, andersen2021adsorption} In the present work we instead fit a single model to all adsorbates, and the different adsorbates are then instead distinguished from each other via adsorbate-specific features such as HOMO/LUMO energy levels. Further information about SISSO is given in Supplementary Section 3.2.

\backmatter

\bmhead{Data availability}
The DFT-calculated adsorption energies and relaxed coordinates of the simple and complex adsorbates databases as well as all calculated features are available at \url{https://github.com/Wenbintum/WWL-GPR} and Zenodo.\cite{wwlgpr2022} Source data for Figures 2, 3, 4 and 5 are available with this manuscript.

\bmhead{Code availability}

The source code of WWL-GPR is publicly available on GitHub at \url{https://github.com/Wenbintum/WWL-GPR} and Zenodo.\cite{wwlgpr2022} We provide pre-defined tasks for tutorial purposes and for reproducing the results presented in this work. The RBF-GPR is implemented with Scikit-learn,\cite{scikit-learn} which is available at \url{https://scikit-learn.org}. The SISSO code \cite{ouyang2018sisso} is available at \url{https://github.com/rouyang2017/SISSO}, and the XGBoost code \cite{chen2016xgboost} is available at \url{https://github.com/dmlc/xgboost}.

\bmhead{Supplementary Information}
Additional details on DFT databases, primary features, machine learning models, and kernel principle component analysis are provided.

\bmhead{Acknowledgments}
The authors gratefully acknowledge support from the Max Planck Computing and Data Facility (MPCDF) and the J{\"u}lich Supercomputing Centre (www.fz-juelich.de/ias/jsc). W.X.\ is grateful for support through the China Scholarship Council (CSC). M.A.\ acknowledges funding from the European Union’s Horizon 2020 research and innovation programme under the Marie Sk\l{}odowska-Curie grant agreement No 754513, the Aarhus University Research Foundation, the Danish National Research Foundation through the Center of Excellence 'InterCat' (Grant agreement no.: DNRF150) and VILLUM FONDEN (grant no.\ 37381).

\bmhead{Author contributions}
W.X.\ carried out the DFT calculations, workflow and ML methods development. K.R.\ and M.A.\ conceived and supervised the project. All authors contributed to analyzing the data and writing the manuscript.

\bmhead{Competing interests}
The authors declare no competing interests.



\clearpage





\begin{thebibliography}{55}
\ifx \bisbn   \undefined \def \bisbn  #1{ISBN #1}\fi
\ifx \binits  \undefined \def \binits#1{#1}\fi
\ifx \bauthor  \undefined \def \bauthor#1{#1}\fi
\ifx \batitle  \undefined \def \batitle#1{#1}\fi
\ifx \bjtitle  \undefined \def \bjtitle#1{#1}\fi
\ifx \bvolume  \undefined \def \bvolume#1{\textbf{#1}}\fi
\ifx \byear  \undefined \def \byear#1{#1}\fi
\ifx \bissue  \undefined \def \bissue#1{#1}\fi
\ifx \bfpage  \undefined \def \bfpage#1{#1}\fi
\ifx \blpage  \undefined \def \blpage #1{#1}\fi
\ifx \burl  \undefined \def \burl#1{\textsf{#1}}\fi
\ifx \doiurl  \undefined \def \doiurl#1{\url{https://doi.org/#1}}\fi
\ifx \betal  \undefined \def \betal{\textit{et al.}}\fi
\ifx \binstitute  \undefined \def \binstitute#1{#1}\fi
\ifx \binstitutionaled  \undefined \def \binstitutionaled#1{#1}\fi
\ifx \bctitle  \undefined \def \bctitle#1{#1}\fi
\ifx \beditor  \undefined \def \beditor#1{#1}\fi
\ifx \bpublisher  \undefined \def \bpublisher#1{#1}\fi
\ifx \bbtitle  \undefined \def \bbtitle#1{#1}\fi
\ifx \bedition  \undefined \def \bedition#1{#1}\fi
\ifx \bseriesno  \undefined \def \bseriesno#1{#1}\fi
\ifx \blocation  \undefined \def \blocation#1{#1}\fi
\ifx \bsertitle  \undefined \def \bsertitle#1{#1}\fi
\ifx \bsnm \undefined \def \bsnm#1{#1}\fi
\ifx \bsuffix \undefined \def \bsuffix#1{#1}\fi
\ifx \bparticle \undefined \def \bparticle#1{#1}\fi
\ifx \barticle \undefined \def \barticle#1{#1}\fi
\bibcommenthead
\ifx \bconfdate \undefined \def \bconfdate #1{#1}\fi
\ifx \botherref \undefined \def \botherref #1{#1}\fi
\ifx \url \undefined \def \url#1{\textsf{#1}}\fi
\ifx \bchapter \undefined \def \bchapter#1{#1}\fi
\ifx \bbook \undefined \def \bbook#1{#1}\fi
\ifx \bcomment \undefined \def \bcomment#1{#1}\fi
\ifx \oauthor \undefined \def \oauthor#1{#1}\fi
\ifx \citeauthoryear \undefined \def \citeauthoryear#1{#1}\fi
\ifx \endbibitem  \undefined \def \endbibitem {}\fi
\ifx \bconflocation  \undefined \def \bconflocation#1{#1}\fi
\ifx \arxivurl  \undefined \def \arxivurl#1{\textsf{#1}}\fi
\csname PreBibitemsHook\endcsname

\bibitem{cao2018mechanistic}
\begin{barticle}
\bauthor{\bsnm{Cao}, \binits{A.}},
\bauthor{\bsnm{Schumann}, \binits{J.}},
\bauthor{\bsnm{Wang}, \binits{T.}},
\bauthor{\bsnm{Zhang}, \binits{L.}},
\bauthor{\bsnm{Xiao}, \binits{J.}},
\bauthor{\bsnm{Bothra}, \binits{P.}},
\bauthor{\bsnm{Liu}, \binits{Y.}},
\bauthor{\bsnm{{Abild-Pedersen}}, \binits{F.}},
\bauthor{\bsnm{N{\o}rskov}, \binits{J.K.}}:
\batitle{Mechanistic insights into the synthesis of higher alcohols from syngas
  on {CuCo} alloys}.
\bjtitle{ACS Catal.}
\bvolume{8}(\bissue{11}),
\bfpage{10148}--\blpage{10155}
(\byear{2018})
\end{barticle}
\endbibitem

\bibitem{chang2021application}
\begin{barticle}
\bauthor{\bsnm{Chang}, \binits{C.}},
\bauthor{\bsnm{Medford}, \binits{A.J.}}:
\batitle{Application of density functional tight binding and machine learning
  to evaluate the stability of biomass intermediates on the {R}h(111) surface}.
\bjtitle{J. Phys. Chem. C}
\bvolume{125}(\bissue{33}),
\bfpage{18210}--\blpage{18216}
(\byear{2021})
\end{barticle}
\endbibitem

\bibitem{wang2020co2}
\begin{botherref}
\oauthor{\bsnm{Wang}, \binits{Z.}},
\oauthor{\bsnm{Li}, \binits{Y.}},
\oauthor{\bsnm{Boes}, \binits{J.}},
\oauthor{\bsnm{Wang}, \binits{Y.}},
\oauthor{\bsnm{Sargent}, \binits{E.}}:
{CO$_2$} electrocatalyst design using graph theory
(2020).
Preprint at \url{https://doi.org/10.21203/rs.3.rs-66715/v1}
\end{botherref}
\endbibitem


\bibitem{norskov2011density}
\begin{barticle}
\bauthor{\bsnm{N{\o}rskov}, \binits{J.K.}},
\bauthor{\bsnm{Abild-Pedersen}, \binits{F.}},
\bauthor{\bsnm{Studt}, \binits{F.}},
\bauthor{\bsnm{Bligaard}, \binits{T.}}:
\batitle{Density functional theory in surface chemistry and catalysis}.
\bjtitle{Proc. Natl. Acad. Sci.}
\bvolume{108}(\bissue{3}),
\bfpage{937}--\blpage{943}
(\byear{2011})
\end{barticle}
\endbibitem


\bibitem{choi2009mechanism}
\begin{barticle}
\bauthor{\bsnm{Choi}, \binits{Y.}},
\bauthor{\bsnm{Liu}, \binits{P.}}:
\batitle{Mechanism of ethanol synthesis from syngas on {R}h (111)}.
\bjtitle{J. Am. Chem. Soc.}
\bvolume{131}(\bissue{36}),
\bfpage{13054}--\blpage{13061}
(\byear{2009})
\end{barticle}
\endbibitem

\bibitem{michel2011c}
\begin{barticle}
\bauthor{\bsnm{Michel}, \binits{C.}},
\bauthor{\bsnm{Auneau}, \binits{F.}},
\bauthor{\bsnm{Delbecq}, \binits{F.}},
\bauthor{\bsnm{Sautet}, \binits{P.}}:
\batitle{C--{H} versus {O}--{H} bond dissociation for alcohols on a {R}h (111)
  surface: A strong assistance from hydrogen bonded neighbors}.
\bjtitle{ACS Catal.}
\bvolume{1}(\bissue{10}),
\bfpage{1430}--\blpage{1440}
(\byear{2011})
\end{barticle}
\endbibitem

\bibitem{filot2015firstprinciplesbased}
\begin{barticle}
\bauthor{\bsnm{Filot}, \binits{I.A.W.}},
\bauthor{\bsnm{Broos}, \binits{R.J.P.}},
\bauthor{\bsnm{{van Rijn}}, \binits{J.P.M.}},
\bauthor{\bsnm{{van Heugten}}, \binits{G.J.H.A.}},
\bauthor{\bsnm{{van Santen}}, \binits{R.A.}},
\bauthor{\bsnm{Hensen}, \binits{E.J.M.}}:
\batitle{First-principles-based microkinetics simulations of synthesis gas
  conversion on a stepped rhodium surface}.
\bjtitle{ACS Catal.}
\bvolume{5}(\bissue{9}),
\bfpage{5453}--\blpage{5467}
(\byear{2015})
\end{barticle}
\endbibitem

\bibitem{gu2020automated}
\begin{barticle}
\bauthor{\bsnm{Gu}, \binits{T.}},
\bauthor{\bsnm{Wang}, \binits{B.}},
\bauthor{\bsnm{Chen}, \binits{S.}},
\bauthor{\bsnm{Yang}, \binits{B.}}:
\batitle{Automated generation and analysis of the complex catalytic reaction
  network of ethanol synthesis from syngas on {Rh}(111)}.
\bjtitle{ACS Catal.}
\bvolume{10}(\bissue{11}),
\bfpage{6346}--\blpage{6355}
(\byear{2020})
\end{barticle}
\endbibitem

\bibitem{tran2018active}
\begin{barticle}
\bauthor{\bsnm{Tran}, \binits{K.}},
\bauthor{\bsnm{Ulissi}, \binits{Z.W.}}:
\batitle{Active learning across intermetallics to guide discovery of
  electrocatalysts for {{CO$_2$}} reduction and {{H$_2$}} evolution}.
\bjtitle{Nat. Catal.}
\bvolume{1}(\bissue{9}),
\bfpage{696}--\blpage{703}
(\byear{2018})
\end{barticle}
\endbibitem

\bibitem{noh2018active}
\begin{barticle}
\bauthor{\bsnm{Noh}, \binits{J.}},
\bauthor{\bsnm{Back}, \binits{S.}},
\bauthor{\bsnm{Kim}, \binits{J.}},
\bauthor{\bsnm{Jung}, \binits{Y.}}:
\batitle{Active learning with non-{\emph{ab initio}} input features toward
  efficient {CO$_2$} reduction catalysts}.
\bjtitle{Chem. Sci.}
\bvolume{9}(\bissue{23}),
\bfpage{5152}--\blpage{5159}
(\byear{2018})
\end{barticle}
\endbibitem

\bibitem{andersen2019scaling}
\begin{barticle}
\bauthor{\bsnm{Andersen}, \binits{M.}},
\bauthor{\bsnm{Levchenko}, \binits{S.V.}},
\bauthor{\bsnm{Scheffler}, \binits{M.}},
\bauthor{\bsnm{Reuter}, \binits{K.}}:
\batitle{Beyond scaling relations for the description of catalytic materials}.
\bjtitle{ACS Catal.}
\bvolume{9}(\bissue{4}),
\bfpage{2752}--\blpage{2759}
(\byear{2019})
\end{barticle}
\endbibitem

\bibitem{wang2021infusing}
\begin{barticle}
\bauthor{\bsnm{Wang}, \binits{S.-H.}},
\bauthor{\bsnm{Pillai}, \binits{H.S.}},
\bauthor{\bsnm{Wang}, \binits{S.}},
\bauthor{\bsnm{Achenie}, \binits{L.E.}},
\bauthor{\bsnm{Xin}, \binits{H.}}:
\batitle{Infusing theory into deep learning for interpretable reactivity
  prediction}.
\bjtitle{Nat. Commun.}
\bvolume{12}(\bissue{1}),
\bfpage{1}--\blpage{9}
(\byear{2021})
\end{barticle}
\endbibitem

\bibitem{fung2021machine}
\begin{barticle}
\bauthor{\bsnm{Fung}, \binits{V.}},
\bauthor{\bsnm{Hu}, \binits{G.}},
\bauthor{\bsnm{Ganesh}, \binits{P.}},
\bauthor{\bsnm{Sumpter}, \binits{B.G.}}:
\batitle{Machine learned features from density of states for accurate
  adsorption energy prediction}.
\bjtitle{Nat. Commun.}
\bvolume{12}(\bissue{1}),
\bfpage{88}
(\byear{2021})
\end{barticle}
\endbibitem



\bibitem{back2019convolutional}
\begin{barticle}
\bauthor{\bsnm{Back}, \binits{S.}},
\bauthor{\bsnm{Yoon}, \binits{J.}},
\bauthor{\bsnm{Tian}, \binits{N.}},
\bauthor{\bsnm{Zhong}, \binits{W.}},
\bauthor{\bsnm{Tran}, \binits{K.}},
\bauthor{\bsnm{Ulissi}, \binits{Z.W.}}:
\batitle{Convolutional neural network of atomic surface structures to predict
  binding energies for high-throughput screening of catalysts}.
\bjtitle{J. Phys. Chem. Lett.}
\bvolume{10}(\bissue{15}),
\bfpage{4401}--\blpage{4408}
(\byear{2019})
\end{barticle}
\endbibitem

\bibitem{gu2020practical}
\begin{barticle}
\bauthor{\bsnm{Gu}, \binits{G.H.}},
\bauthor{\bsnm{Noh}, \binits{J.}},
\bauthor{\bsnm{Kim}, \binits{S.}},
\bauthor{\bsnm{Back}, \binits{S.}},
\bauthor{\bsnm{Ulissi}, \binits{Z.}},
\bauthor{\bsnm{Jung}, \binits{Y.}}:
\batitle{Practical deep-learning representation for fast heterogeneous catalyst
  screening}.
\bjtitle{J. Phys. Chem. Lett.}
\bvolume{11}(\bissue{9}),
\bfpage{3185}--\blpage{3191}
(\byear{2020})
\end{barticle}
\endbibitem


\bibitem{chanussot2021open}
\begin{barticle}
\bauthor{\bsnm{Chanussot}, \binits{L.}},
\bauthor{\bsnm{Das}, \binits{A.}},
\bauthor{\bsnm{Goyal}, \binits{S.}},
\bauthor{\bsnm{Lavril}, \binits{T.}},
\bauthor{\bsnm{Shuaibi}, \binits{M.}},
\bauthor{\bsnm{Riviere}, \binits{M.}},
\bauthor{\bsnm{Tran}, \binits{K.}},
\bauthor{\bsnm{Heras-Domingo}, \binits{J.}},
\bauthor{\bsnm{Ho}, \binits{C.}},
\bauthor{\bsnm{Hu}, \binits{W.}},
\bauthor{\bsnm{Palizhati}, \binits{A.}},
\bauthor{\bsnm{Sriram}, \binits{A.}},
\bauthor{\bsnm{Wood}, \binits{B.}},
\bauthor{\bsnm{Yoon}, \binits{J.}},
\bauthor{\bsnm{Parikh}, \binits{D.}},
\bauthor{\bsnm{Zitnick}, \binits{C.L.}},
\bauthor{\bsnm{Ulissi}, \binits{Z.}}:
\batitle{Open {Catalyst} 2020 {(OC20)} dataset and community challenges}.
\bjtitle{ACS Catal.}
\bvolume{11}(\bissue{10}),
\bfpage{6059}--\blpage{6072}
(\byear{2021})
\end{barticle}
\endbibitem

\bibitem{togninalli2019wasserstein}
\begin{botherref}
\oauthor{\bsnm{Togninalli}, \binits{M.}},
\oauthor{\bsnm{Ghisu}, \binits{E.}},
\oauthor{\bsnm{Llinares-L{\'o}pez}, \binits{F.}},
\oauthor{\bsnm{Rieck}, \binits{B.}},
\oauthor{\bsnm{Borgwardt}, \binits{K.}}:
Wasserstein {{Weisfeiler}}-{{Lehman Graph Kernels}}.
Adv Neural Inf Process Syst.
\textbf{32}
(2019)
\end{botherref}
\endbibitem

\bibitem{ouyang2018sisso}
\begin{barticle}
\bauthor{\bsnm{Ouyang}, \binits{R.}},
\bauthor{\bsnm{Curtarolo}, \binits{S.}},
\bauthor{\bsnm{Ahmetcik}, \binits{E.}},
\bauthor{\bsnm{Scheffler}, \binits{M.}},
\bauthor{\bsnm{Ghiringhelli}, \binits{L.M.}}:
\batitle{{SISSO}: A compressed-sensing method for identifying the best
  low-dimensional descriptor in an immensity of offered candidates}.
\bjtitle{Phys. Rev. Mater.}
\bvolume{2}(\bissue{8}),
\bfpage{083802}
(\byear{2018})
\end{barticle}
\endbibitem

\bibitem{ouyang2019simultaneous}
\begin{barticle}
\bauthor{\bsnm{Ouyang}, \binits{R.}},
\bauthor{\bsnm{Ahmetcik}, \binits{E.}},
\bauthor{\bsnm{Carbogno}, \binits{C.}},
\bauthor{\bsnm{Scheffler}, \binits{M.}},
\bauthor{\bsnm{Ghiringhelli}, \binits{L.M.}}:
\batitle{Simultaneous learning of several materials properties from incomplete
  databases with multi-task {SISSO}}.
\bjtitle{J. Phys.: Mater.}
\bvolume{2}(\bissue{2}),
\bfpage{024002}
(\byear{2019})
\end{barticle}
\endbibitem

\bibitem{chen2016xgboost}
\begin{bchapter}
\bauthor{\bsnm{Chen}, \binits{T.}},
\bauthor{\bsnm{Guestrin}, \binits{C.}}:
\bctitle{Xgboost: A scalable tree boosting system}.
In: \bbtitle{Proceedings of the 22nd Acm Sigkdd International Conference on
  Knowledge Discovery and Data Mining},
pp. \bfpage{785}--\blpage{794}
(\byear{2016})
\end{bchapter}
\endbibitem

\bibitem{medford2014activity}
\begin{barticle}
\bauthor{\bsnm{Medford}, \binits{A.J.}},
\bauthor{\bsnm{Lausche}, \binits{A.C.}},
\bauthor{\bsnm{Abild-Pedersen}, \binits{F.}},
\bauthor{\bsnm{Temel}, \binits{B.}},
\bauthor{\bsnm{Schj{\o}dt}, \binits{N.C.}},
\bauthor{\bsnm{N{\o}rskov}, \binits{J.K.}},
\bauthor{\bsnm{Studt}, \binits{F.}}:
\batitle{Activity and selectivity trends in synthesis gas conversion to higher
  alcohols}.
\bjtitle{Top. Catal.}
\bvolume{57}(\bissue{1}),
\bfpage{135}--\blpage{142}
(\byear{2014})
\end{barticle}
\endbibitem

\bibitem{schumann2018selectivity}
\begin{barticle}
\bauthor{\bsnm{Schumann}, \binits{J.}},
\bauthor{\bsnm{Medford}, \binits{A.J.}},
\bauthor{\bsnm{Yoo}, \binits{J.S.}},
\bauthor{\bsnm{Zhao}, \binits{Z.-J.}},
\bauthor{\bsnm{Bothra}, \binits{P.}},
\bauthor{\bsnm{Cao}, \binits{A.}},
\bauthor{\bsnm{Studt}, \binits{F.}},
\bauthor{\bsnm{Abild-Pedersen}, \binits{F.}},
\bauthor{\bsnm{Nørskov}, \binits{J.K.}}:
\batitle{Selectivity of synthesis gas conversion to c2+ oxygenates on fcc(111)
  transition-metal surfaces}.
\bjtitle{ACS Catal.}
\bvolume{8}(\bissue{4}),
\bfpage{3447}--\blpage{3453}
(\byear{2018})
\end{barticle}
\endbibitem


\bibitem{deimel2020active}
\begin{barticle}
\bauthor{\bsnm{Deimel}, \binits{M.}},
\bauthor{\bsnm{Reuter}, \binits{K.}},
\bauthor{\bsnm{Andersen}, \binits{M.}}:
\batitle{Active site representation in first-principles microkinetic models:
  Data-enhanced computational screening for improved methanation catalysts}.
\bjtitle{ACS Catal.}
\bvolume{10}(\bissue{22}),
\bfpage{13729}--\blpage{13736}
(\byear{2020})
\end{barticle}
\endbibitem

\bibitem{deringer2019machine}
\begin{barticle}
\bauthor{\bsnm{Deringer}, \binits{V.L.}},
\bauthor{\bsnm{Caro}, \binits{M.A.}},
\bauthor{\bsnm{Cs{\'a}nyi}, \binits{G.}}:
\batitle{Machine learning interatomic potentials as emerging tools for
  materials science}.
\bjtitle{Adv. Mater.}
\bvolume{31}(\bissue{46}),
\bfpage{1902765}
(\byear{2019})
\end{barticle}
\endbibitem

\bibitem{gasteiger_gemnet_2021}
\begin{bchapter}
\bauthor{\bsnm{Gasteiger}, \binits{J.}},
\bauthor{\bsnm{Becker}, \binits{F.}},
\bauthor{\bsnm{G{\"u}nnemann}, \binits{S.}}:
\bctitle{Gemnet: Universal directional graph neural networks for molecules}.
In: \bbtitle{Conference on Neural Information Processing Systems (NeurIPS)}
(\byear{2021})
\end{bchapter}
\endbibitem

\bibitem{wen2021bondnet}
\begin{barticle}
\bauthor{\bsnm{Wen}, \binits{M.}},
\bauthor{\bsnm{Blau}, \binits{S.M.}},
\bauthor{\bsnm{{Spotte-Smith}}, \binits{E.W.C.}},
\bauthor{\bsnm{Dwaraknath}, \binits{S.}},
\bauthor{\bsnm{Persson}, \binits{K.A.}}:
\batitle{{{BonDNet}}: A graph neural network for the prediction of bond
  dissociation energies for charged molecules}.
\bjtitle{Chem. Sci.}
\bvolume{12}(\bissue{5}),
\bfpage{1858}--\blpage{1868}
(\byear{2021})
\end{barticle}
\endbibitem

\bibitem{tang2019prediction}
\begin{barticle}
\bauthor{\bsnm{Tang}, \binits{Y.-H.}},
\bauthor{\bsnm{{de Jong}}, \binits{W.A.}}:
\batitle{Prediction of atomization energy using graph kernel and active
  learning}.
\bjtitle{J. Chem. Phys.}
\bvolume{150}(\bissue{4}),
\bfpage{044107}
(\byear{2019})
\end{barticle}
\endbibitem

\bibitem{xie2018crystal}
\begin{barticle}
\bauthor{\bsnm{Xie}, \binits{T.}},
\bauthor{\bsnm{Grossman}, \binits{J.C.}}:
\batitle{Crystal graph convolutional neural networks for an accurate and
  interpretable prediction of material properties}.
\bjtitle{Phys. Rev. Lett.}
\bvolume{120}(\bissue{14}),
\bfpage{145301}
(\byear{2018})
\end{barticle}
\endbibitem

\bibitem{montoya2017highthroughput}
\begin{barticle}
\bauthor{\bsnm{Montoya}, \binits{J.H.}},
\bauthor{\bsnm{Persson}, \binits{K.A.}}:
\batitle{A high-throughput framework for determining adsorption energies on
  solid surfaces}.
\bjtitle{Npj Comput. Mater.}
\bvolume{3}(\bissue{1}),
\bfpage{14}
(\byear{2017})
\end{barticle}
\endbibitem

\bibitem{boes2019graph}
\begin{barticle}
\bauthor{\bsnm{Boes}, \binits{J.R.}},
\bauthor{\bsnm{Mamun}, \binits{O.}},
\bauthor{\bsnm{Winther}, \binits{K.}},
\bauthor{\bsnm{Bligaard}, \binits{T.}}:
\batitle{Graph theory approach to high-throughput surface adsorption structure
  generation}.
\bjtitle{J. Phys. Chem. A}
\bvolume{123}(\bissue{11}),
\bfpage{2281}--\blpage{2285}
(\byear{2019})
\end{barticle}
\endbibitem

\bibitem{deshpande2020graph}
\begin{barticle}
\bauthor{\bsnm{Deshpande}, \binits{S.}},
\bauthor{\bsnm{Maxson}, \binits{T.}},
\bauthor{\bsnm{Greeley}, \binits{J.}}:
\batitle{Graph theory approach to determine configurations of multidentate and
  high coverage adsorbates for heterogeneous catalysis}.
\bjtitle{Npj Comput. Mater.}
\bvolume{6}(\bissue{1}),
\bfpage{79}
(\byear{2020})
\end{barticle}
\endbibitem

\bibitem{xu2020data}
\begin{barticle}
\bauthor{\bsnm{Xu}, \binits{W.}},
\bauthor{\bsnm{Andersen}, \binits{M.}},
\bauthor{\bsnm{Reuter}, \binits{K.}}:
\batitle{Data-driven descriptor engineering and refined scaling relations for
  predicting transition metal oxide reactivity}.
\bjtitle{ACS Catal.}
\bvolume{11}(\bissue{2}),
\bfpage{734}--\blpage{742}
(\byear{2020})
\end{barticle}
\endbibitem

\bibitem{rupp2015machine}
\begin{barticle}
\bauthor{\bsnm{Rupp}, \binits{M.}}:
\batitle{Machine learning for quantum mechanics in a nutshell}.
\bjtitle{Int. J. Quantum Chem.}
\bvolume{115}(\bissue{16}),
\bfpage{1058}--\blpage{1073}
(\byear{2015})
\end{barticle}
\endbibitem

\bibitem{deringer2021gaussian}
\begin{barticle}
\bauthor{\bsnm{Deringer}, \binits{V.L.}},
\bauthor{\bsnm{Bart{\'o}k}, \binits{A.P.}},
\bauthor{\bsnm{Bernstein}, \binits{N.}},
\bauthor{\bsnm{Wilkins}, \binits{D.M.}},
\bauthor{\bsnm{Ceriotti}, \binits{M.}},
\bauthor{\bsnm{Cs{\'a}nyi}, \binits{G.}}:
\batitle{Gaussian process regression for materials and molecules}.
\bjtitle{Chem. Rev.}
\bvolume{121}(\bissue{16}),
\bfpage{10073}--\blpage{10141}
(\byear{2021})
\end{barticle}
\endbibitem

\bibitem{bruix2019first}
\begin{barticle}
\bauthor{\bsnm{Bruix}, \binits{A.}},
\bauthor{\bsnm{Margraf}, \binits{J.T.}},
\bauthor{\bsnm{Andersen}, \binits{M.}},
\bauthor{\bsnm{Reuter}, \binits{K.}}:
\batitle{First-principles-based multiscale modelling of heterogeneous
  catalysis}.
\bjtitle{Nat. Catal.}
\bvolume{2}(\bissue{8}),
\bfpage{659}--\blpage{670}
(\byear{2019})
\end{barticle}
\endbibitem

\bibitem{meskine2009examination}
\begin{barticle}
\bauthor{\bsnm{Meskine}, \binits{H.}},
\bauthor{\bsnm{Matera}, \binits{S.}},
\bauthor{\bsnm{Scheffler}, \binits{M.}},
\bauthor{\bsnm{Reuter}, \binits{K.}},
\bauthor{\bsnm{Metiu}, \binits{H.}}:
\batitle{Examination of the concept of degree of rate control by
  first-principles kinetic monte carlo simulations}.
\bjtitle{Surf. Sci.}
\bvolume{603}(\bissue{10-12}),
\bfpage{1724}--\blpage{1730}
(\byear{2009})
\end{barticle}
\endbibitem

\bibitem{medford2014assessing}
\begin{barticle}
\bauthor{\bsnm{Medford}, \binits{A.J.}},
\bauthor{\bsnm{Wellendorff}, \binits{J.}},
\bauthor{\bsnm{Vojvodic}, \binits{A.}},
\bauthor{\bsnm{Studt}, \binits{F.}},
\bauthor{\bsnm{Abild-Pedersen}, \binits{F.}},
\bauthor{\bsnm{Jacobsen}, \binits{K.W.}},
\bauthor{\bsnm{Bligaard}, \binits{T.}},
\bauthor{\bsnm{N{\o}rskov}, \binits{J.K.}}:
\batitle{Assessing the reliability of calculated catalytic ammonia synthesis
  rates}.
\bjtitle{Science}
\bvolume{345}(\bissue{6193}),
\bfpage{197}--\blpage{200}
(\byear{2014})
\end{barticle}
\endbibitem

\bibitem{sutton2016effects}
\begin{barticle}
\bauthor{\bsnm{Sutton}, \binits{J.E.}},
\bauthor{\bsnm{Guo}, \binits{W.}},
\bauthor{\bsnm{Katsoulakis}, \binits{M.A.}},
\bauthor{\bsnm{Vlachos}, \binits{D.G.}}:
\batitle{Effects of correlated parameters and uncertainty in
  electronic-structure-based chemical kinetic modelling}.
\bjtitle{Nat. Chem.}
\bvolume{8}(\bissue{4}),
\bfpage{331}
(\byear{2016})
\end{barticle}
\endbibitem

\bibitem{dopking2017error}
\begin{barticle}
\bauthor{\bsnm{D{\"o}pking}, \binits{S.}},
\bauthor{\bsnm{Matera}, \binits{S.}}:
\batitle{Error propagation in first-principles kinetic monte carlo simulation}.
\bjtitle{Chem. Phys. Lett.}
\bvolume{674},
\bfpage{28}--\blpage{32}
(\byear{2017})
\end{barticle}
\endbibitem

\bibitem{flores2020active}
\begin{barticle}
\bauthor{\bsnm{Flores}, \binits{R.A.}},
\bauthor{\bsnm{Paolucci}, \binits{C.}},
\bauthor{\bsnm{Winther}, \binits{K.T.}},
\bauthor{\bsnm{Jain}, \binits{A.}},
\bauthor{\bsnm{Torres}, \binits{J.A.G.}},
\bauthor{\bsnm{Aykol}, \binits{M.}},
\bauthor{\bsnm{Montoya}, \binits{J.}},
\bauthor{\bsnm{N{\o}rskov}, \binits{J.K.}},
\bauthor{\bsnm{Bajdich}, \binits{M.}},
\bauthor{\bsnm{Bligaard}, \binits{T.}}:
\batitle{Active learning accelerated discovery of stable iridium oxide
  polymorphs for the oxygen evolution reaction}.
\bjtitle{Chem. Mater.}
\bvolume{32}(\bissue{13}),
\bfpage{5854}--\blpage{5863}
(\byear{2020})
\end{barticle}
\endbibitem

\bibitem{kunkel2021active}
\begin{barticle}
\bauthor{\bsnm{Kunkel}, \binits{C.}},
\bauthor{\bsnm{Margraf}, \binits{J.T.}},
\bauthor{\bsnm{Chen}, \binits{K.}},
\bauthor{\bsnm{Oberhofer}, \binits{H.}},
\bauthor{\bsnm{Reuter}, \binits{K.}}:
\batitle{Active discovery of organic semiconductors}.
\bjtitle{Nat. Commun.}
\bvolume{12}(\bissue{1}),
\bfpage{1}--\blpage{11}
(\byear{2021})
\end{barticle}
\endbibitem

\bibitem{tran2020methods}
\begin{barticle}
\bauthor{\bsnm{Tran}, \binits{K.}},
\bauthor{\bsnm{Neiswanger}, \binits{W.}},
\bauthor{\bsnm{Yoon}, \binits{J.}},
\bauthor{\bsnm{Zhang}, \binits{Q.}},
\bauthor{\bsnm{Xing}, \binits{E.}},
\bauthor{\bsnm{Ulissi}, \binits{Z.W.}}:
\batitle{Methods for comparing uncertainty quantifications for material
  property predictions}.
\bjtitle{Mach. Learn.: Sci. Technol.}
\bvolume{1}(\bissue{2}),
\bfpage{025006}
(\byear{2020})
\end{barticle}
\endbibitem


\bibitem{palmer2022calibration}
\begin{barticle}
\bauthor{\bsnm{Palmer}, \binits{G.}},
\bauthor{\bsnm{Du}, \binits{S.}},
\bauthor{\bsnm{Politowicz}, \binits{A.}},
\bauthor{\bsnm{Emory}, \binits{J.P.}},
\bauthor{\bsnm{Yang}, \binits{X.}},
\bauthor{\bsnm{Gautam}, \binits{A.}},
\bauthor{\bsnm{Gupta}, \binits{G.}},
\bauthor{\bsnm{Li}, \binits{Z.}},
\bauthor{\bsnm{Jacobs}, \binits{R.}},
\bauthor{\bsnm{Morgan}, \binits{D.}}:
\batitle{Calibration after bootstrap for accurate uncertainty quantification in
  regression models}.
\bjtitle{npj Computational Materials}
\bvolume{8}(\bissue{1}),
\bfpage{1}--\blpage{9}
(\byear{2022})
\end{barticle}
\endbibitem


\bibitem{kuleshov2018accurate}
\begin{bchapter}
\bauthor{\bsnm{Kuleshov}, \binits{V.}},
\bauthor{\bsnm{Fenner}, \binits{N.}},
\bauthor{\bsnm{Ermon}, \binits{S.}}:
\bctitle{Accurate uncertainties for deep learning using calibrated regression}.
In: \bbtitle{International Conference on Machine Learning},
pp. \bfpage{2796}--\blpage{2804}
(\byear{2018}).
\bcomment{PMLR}
\end{bchapter}
\endbibitem



\bibitem{Giannozzi_2009}
\begin{barticle}
\bauthor{\bsnm{Giannozzi}, \binits{P.}},
\bauthor{\bsnm{Baroni}, \binits{S.}},
\bauthor{\bsnm{Bonini}, \binits{N.}},
\bauthor{\bsnm{Calandra}, \binits{M.}},
\bauthor{\bsnm{Car}, \binits{R.}},
\bauthor{\bsnm{Cavazzoni}, \binits{C.}},
\bauthor{\bsnm{Ceresoli}, \binits{D.}},
\bauthor{\bsnm{Chiarotti}, \binits{G.L.}},
\bauthor{\bsnm{Cococcioni}, \binits{M.}},
\bauthor{\bsnm{Dabo}, \binits{I.}},
\bauthor{\bsnm{Corso}, \binits{A.D.}},
\bauthor{\bparticle{de} \bsnm{Gironcoli}, \binits{S.}},
\bauthor{\bsnm{Fabris}, \binits{S.}},
\bauthor{\bsnm{Fratesi}, \binits{G.}},
\bauthor{\bsnm{Gebauer}, \binits{R.}},
\bauthor{\bsnm{Gerstmann}, \binits{U.}},
\bauthor{\bsnm{Gougoussis}, \binits{C.}},
\bauthor{\bsnm{Kokalj}, \binits{A.}},
\bauthor{\bsnm{Lazzeri}, \binits{M.}},
\bauthor{\bsnm{Martin-Samos}, \binits{L.}},
\bauthor{\bsnm{Marzari}, \binits{N.}},
\bauthor{\bsnm{Mauri}, \binits{F.}},
\bauthor{\bsnm{Mazzarello}, \binits{R.}},
\bauthor{\bsnm{Paolini}, \binits{S.}},
\bauthor{\bsnm{Pasquarello}, \binits{A.}},
\bauthor{\bsnm{Paulatto}, \binits{L.}},
\bauthor{\bsnm{Sbraccia}, \binits{C.}},
\bauthor{\bsnm{Scandolo}, \binits{S.}},
\bauthor{\bsnm{Sclauzero}, \binits{G.}},
\bauthor{\bsnm{Seitsonen}, \binits{A.P.}},
\bauthor{\bsnm{Smogunov}, \binits{A.}},
\bauthor{\bsnm{Umari}, \binits{P.}},
\bauthor{\bsnm{Wentzcovitch}, \binits{R.M.}}:
\batitle{{QUANTUM} {ESPRESSO}: a modular and open-source software project for
  quantum simulations of materials}.
\bjtitle{J. Phys.: Condens. Matter}
\bvolume{21}(\bissue{39}),
\bfpage{395502}
(\byear{2009})
\end{barticle}
\endbibitem

\bibitem{wellendorff2012density}
\begin{barticle}
\bauthor{\bsnm{Wellendorff}, \binits{J.}},
\bauthor{\bsnm{Lundgaard}, \binits{K.T.}},
\bauthor{\bsnm{M{\o}gelh{\o}j}, \binits{A.}},
\bauthor{\bsnm{Petzold}, \binits{V.}},
\bauthor{\bsnm{Landis}, \binits{D.D.}},
\bauthor{\bsnm{N{\o}rskov}, \binits{J.K.}},
\bauthor{\bsnm{Bligaard}, \binits{T.}},
\bauthor{\bsnm{Jacobsen}, \binits{K.W.}}:
\batitle{Density functionals for surface science: Exchange-correlation model
  development with bayesian error estimation}.
\bjtitle{Phys. Rev. B}
\bvolume{85}(\bissue{23}),
\bfpage{235149}
(\byear{2012})
\end{barticle}
\endbibitem

\bibitem{huber2020aiida}
\begin{barticle}
\bauthor{\bsnm{Huber}, \binits{S.P.}},
\bauthor{\bsnm{Zoupanos}, \binits{S.}},
\bauthor{\bsnm{Uhrin}, \binits{M.}},
\bauthor{\bsnm{Talirz}, \binits{L.}},
\bauthor{\bsnm{Kahle}, \binits{L.}},
\bauthor{\bsnm{H{\"a}uselmann}, \binits{R.}},
\bauthor{\bsnm{Gresch}, \binits{D.}},
\bauthor{\bsnm{M{\"u}ller}, \binits{T.}},
\bauthor{\bsnm{Yakutovich}, \binits{A.V.}},
\bauthor{\bsnm{Andersen}, \binits{C.W.}},
\bauthor{\bsnm{Ramirez}, \binits{F.F.}},
\bauthor{\bsnm{Adorf}, \binits{C.S.}},
\bauthor{\bsnm{Gargiulo}, \binits{F.}},
\bauthor{\bsnm{Kumbhar}, \binits{S.}},
\bauthor{\bsnm{Passaro}, \binits{E.}},
\bauthor{\bsnm{Johnston}, \binits{C.}},
\bauthor{\bsnm{Merkys}, \binits{A.}},
\bauthor{\bsnm{Cepellotti}, \binits{A.}},
\bauthor{\bsnm{Mounet}, \binits{N.}},
\bauthor{\bsnm{Marzari}, \binits{N.}},
\bauthor{\bsnm{Kozinsky}, \binits{B.}},
\bauthor{\bsnm{Pizzi}, \binits{G.}}:
\batitle{{{AiiDA}} 1.0, a scalable computational infrastructure for automated
  reproducible workflows and data provenance}.
\bjtitle{Sci. Data}
\bvolume{7}(\bissue{1}),
\bfpage{300}
(\byear{2020})
\end{barticle}
\endbibitem

\bibitem{wwlgpr2022}
\begin{botherref}
\oauthor{\bsnm{Xu}, \binits{W.}},
\oauthor{\bsnm{Reuter}, \binits{K.}},
\oauthor{\bsnm{Andersen}, \binits{M.}}:
Predicting binding motifs of complex adsorbates using machine learning with a
  physics-inspired graph representation.
Zenodo.
\url{https://doi.org/10.5281/zenodo.6640198}
(2022)
\end{botherref}
\endbibitem


\bibitem{Hjorth_Larsen_2017}
\begin{barticle}
\bauthor{\bsnm{Larsen}, \binits{A.H.}},
\bauthor{\bsnm{Mortensen}, \binits{J.J.}},
\bauthor{\bsnm{Blomqvist}, \binits{J.}},
\bauthor{\bsnm{Castelli}, \binits{I.E.}},
\bauthor{\bsnm{Christensen}, \binits{R.}},
\bauthor{\bsnm{Du{\l}ak}, \binits{M.}},
\bauthor{\bsnm{Friis}, \binits{J.}},
\bauthor{\bsnm{Groves}, \binits{M.N.}},
\bauthor{\bsnm{Hammer}, \binits{B.}},
\bauthor{\bsnm{Hargus}, \binits{C.}},
\bauthor{\bsnm{Hermes}, \binits{E.D.}},
\bauthor{\bsnm{Jennings}, \binits{P.C.}},
\bauthor{\bsnm{Jensen}, \binits{P.B.}},
\bauthor{\bsnm{Kermode}, \binits{J.}},
\bauthor{\bsnm{Kitchin}, \binits{J.R.}},
\bauthor{\bsnm{Kolsbjerg}, \binits{E.L.}},
\bauthor{\bsnm{Kubal}, \binits{J.}},
\bauthor{\bsnm{Kaasbjerg}, \binits{K.}},
\bauthor{\bsnm{Lysgaard}, \binits{S.}},
\bauthor{\bsnm{Maronsson}, \binits{J.B.}},
\bauthor{\bsnm{Maxson}, \binits{T.}},
\bauthor{\bsnm{Olsen}, \binits{T.}},
\bauthor{\bsnm{Pastewka}, \binits{L.}},
\bauthor{\bsnm{Peterson}, \binits{A.}},
\bauthor{\bsnm{Rostgaard}, \binits{C.}},
\bauthor{\bsnm{Schi{\o}tz}, \binits{J.}},
\bauthor{\bsnm{Schütt}, \binits{O.}},
\bauthor{\bsnm{Strange}, \binits{M.}},
\bauthor{\bsnm{Thygesen}, \binits{K.S.}},
\bauthor{\bsnm{Vegge}, \binits{T.}},
\bauthor{\bsnm{Vilhelmsen}, \binits{L.}},
\bauthor{\bsnm{Walter}, \binits{M.}},
\bauthor{\bsnm{Zeng}, \binits{Z.}},
\bauthor{\bsnm{Jacobsen}, \binits{K.W.}}:
\batitle{The atomic simulation environment{\textemdash}a python library for
  working with atoms}.
\bjtitle{J. Phys.: Condens. Matter}
\bvolume{29}(\bissue{27}),
\bfpage{273002}
(\byear{2017})
\end{barticle}
\endbibitem

\bibitem{dal2014pseudopotentials}
\begin{barticle}
\bauthor{\bsnm{Dal~Corso}, \binits{A.}}:
\batitle{Pseudopotentials periodic table: From {H} to {Pu}}.
\bjtitle{Comput. Mater. Sci.}
\bvolume{95},
\bfpage{337}--\blpage{350}
(\byear{2014})
\end{barticle}
\endbibitem

\bibitem{garrity2014pseudopotentials}
\begin{barticle}
\bauthor{\bsnm{Garrity}, \binits{K.F.}},
\bauthor{\bsnm{Bennett}, \binits{J.W.}},
\bauthor{\bsnm{Rabe}, \binits{K.M.}},
\bauthor{\bsnm{Vanderbilt}, \binits{D.}}:
\batitle{Pseudopotentials for high-throughput {DFT} calculations}.
\bjtitle{Comput. Mater. Sci.}
\bvolume{81},
\bfpage{446}--\blpage{452}
(\byear{2014})
\end{barticle}
\endbibitem

\bibitem{bartok2013representing}
\begin{barticle}
\bauthor{\bsnm{Bart{\'o}k}, \binits{A.P.}},
\bauthor{\bsnm{Kondor}, \binits{R.}},
\bauthor{\bsnm{Cs{\'a}nyi}, \binits{G.}}:
\batitle{On representing chemical environments}.
\bjtitle{Phys. Rev. B.}
\bvolume{87}(\bissue{18}),
\bfpage{184115}
(\byear{2013})
\end{barticle}
\endbibitem

\bibitem{Esterhuizen2020}
\begin{barticle}
\bauthor{\bsnm{Esterhuizen}, \binits{J.A.}},
\bauthor{\bsnm{Goldsmith}, \binits{B.R.}},
\bauthor{\bsnm{Linic}, \binits{S.}}:
\batitle{Theory-guided machine learning finds geometric structure-property
  relationships for chemisorption on subsurface alloys}.
\bjtitle{Chem}
\bvolume{6}(\bissue{11}),
\bfpage{3100}--\blpage{3117}
(\byear{2020})
\end{barticle}
\endbibitem

\bibitem{andersen2021adsorption}
\begin{barticle}
\bauthor{\bsnm{Andersen}, \binits{M.}},
\bauthor{\bsnm{Reuter}, \binits{K.}}:
\batitle{Adsorption enthalpies for catalysis modeling through machine-learned
  descriptors}.
\bjtitle{Acc. Chem. Res.}
\bvolume{54}(\bissue{12}),
\bfpage{2741}--\blpage{2749}
(\byear{2021})
\end{barticle}
\endbibitem



\bibitem{scikit-learn}
\begin{barticle}
\bauthor{\bsnm{Pedregosa}, \binits{F.}},
\bauthor{\bsnm{Varoquaux}, \binits{G.}},
\bauthor{\bsnm{Gramfort}, \binits{A.}},
\bauthor{\bsnm{Michel}, \binits{V.}},
\bauthor{\bsnm{Thirion}, \binits{B.}},
\bauthor{\bsnm{Grisel}, \binits{O.}},
\bauthor{\bsnm{Blondel}, \binits{M.}},
\bauthor{\bsnm{Prettenhofer}, \binits{P.}},
\bauthor{\bsnm{Weiss}, \binits{R.}},
\bauthor{\bsnm{Dubourg}, \binits{V.}},
\bauthor{\bsnm{Vanderplas}, \binits{J.}},
\bauthor{\bsnm{Passos}, \binits{A.}},
\bauthor{\bsnm{Cournapeau}, \binits{D.}},
\bauthor{\bsnm{Brucher}, \binits{M.}},
\bauthor{\bsnm{Perrot}, \binits{M.}},
\bauthor{\bsnm{Duchesnay}, \binits{E.}}:
\batitle{Scikit-learn: machine learning in python}.
\bjtitle{J. Mach. Learn. Res.}
\bvolume{12},
\bfpage{2825}--\blpage{2830}
(\byear{2011})
\end{barticle}
\endbibitem

\end{thebibliography}

\end{document}